\documentclass[12pt]{article}
\usepackage[dvips]{graphicx}

\newcommand{\be}{\begin{equation}}
\newcommand{\ee}{\end{equation}}
\begin{document}
\baselineskip=18pt
\pagestyle{empty}
\setcounter{equation}{0}
\setcounter{page}{1}

\title{Underlying Determinism, Stationary Phase \\and Quantum Mechanics}% Force line breaks with \\
\author{R. Fukuda \protect\\
Department of physics,
Faculty of Science and Technology,
 Keio University}
\date{(29 November 2007)}

\maketitle

%\begin{document}
\begin{center}
\begin{large}
{\bf Abstract}
\end{large}
\end{center}

\vspace{0.3cm}
\noindent
In a newly introduced time scale $\tau$,  
 much smaller than the usual $t$, 
 any object is assumed to be a point-like particle, having 
a definite position. It fluctuates without
 dynamics and the wave function $\Psi$ is defined by averaging the square root of the density.   
 In $t$-scale, the Schr$\ddot{\rm o}$dinger equation holds and  
 for a macrovariable just a classical path is picked up as a peak of $\Psi$ by   
 the stationary phase, which is the observable signal. In the measuring process, 
the stationary phase branches into many but one     
 branch is selected by underlying determinism, 
leading to the correct detection probability. 

\vspace{0.5cm}
\begin{center}
{\bf Introduction}
\end{center}
\vspace{0.2cm}

Observational problem in quantum mechanics has a long history of debates. 
The crucial role of the docoherence in the measurement    
 has been widely discussed\cite{Zurek,Schlosshauser,
Zeh,Omnes,Ghose}. Also  
 the dynamical reduction 
model has actually been constructed\cite{Rimini,Pearle,Leggett,Bassi,BassiSalvetti,ABassi}.
Guided by a sudden change in the detection process 
from the wave to the particle picture, 
producing sample dependent random signals,  
 we assume in this letter that an object is a point-like particle and 
has a fluctuating but definite position 
 $q(\tau)$ (in one dimension, for simplicity) 
 in a new time scale $\tau$. 
$q(\tau)$ just fluctuates uniformly without any dynamics and    
  is not observable. The usual variable $t$ is a coarse grained version 
of $\tau$ and the wave function is defined by summing up   
 coarse grained paths
 in the form of the ``square-root " of the density, 
which accompanies the phase. Coherence in $t$ scale is controlled by this phase.  
  For a  macrovariable, 
the stationary phase mechanism works. It achieves both the complete 
construction and destruction of the coherence, 
selecting  
 a deterministic trajectory of classical type 
 as a peak of the wave function, which is the only signal of the observation. 
The detection apparatus realizes the branching 
 of the stationary phase  
 but due to underlying determinism, one branch is chosen by chance for one sample and  
the desired probability rule is obtained. 
As opposed to Ref.\cite{Rimini}, the Schr$\ddot{\rm o}$dinger equation (SE)  
 holds without any modifications for both micro and macrovariables. 
Our theory is applicable to an isolated system and is 
totally different from that based on the environment\cite{Zurek}, and also   
 from the hidden-variable theory\cite{Bell,Genovese}.

\vspace{0.3cm}
\begin{center}
{\bf Time scale}
\end{center}
\vspace{0.2cm}

Discretized time is used as 
 $t_n =t_0 -n\Delta t$, $\tau_i =\tau_0 -i\Delta \tau$ ($n,i \geq 0$, $t_0 \equiv t$) and write $\Delta t/\Delta \tau=M$. 
Define the interval $D_{t_n}$ 
which contains $M$ points $\tau_{m_n +nM}$, $(m_n =0,\pm 1,\pm 2,\cdots, \pm M/2)$. 
The center position of $D_{t_n}$ is $\tau_{nM}\equiv t_n$ and 
$\Delta \tau$ goes to zero before $\Delta t\rightarrow 0$ assuring $M\rightarrow\infty$. 
Take one sample $q(\tau)$ and write $q(\tau_{m_n +nM}) \equiv q_{m_n}$.  
For any fixed $n$, $M$ points of $q_{m_n}$ are assumed to 
distribute uniformly over all space as $M\rightarrow\infty$. 
They are mutually exclusive by determinism. 
Selecting one $m_n$ in every $D_{t_n}$,
a coarse grained path 
${\mathcal{P}}\equiv (q_{m_{0}},q_{m_{1}},q_{m_{2}},\cdots)$ 
is introduced. 
\vspace{0.3cm}
\begin{center}
{\bf Wave function $\Psi$}
\end{center}
\vspace{0.2cm}

Consider the density at $t$,  
 $\rho(x,t)=\delta(x-q_{m_0})$. Being 
   positive definite, 
  it can be written as $\psi^{\dagger}\psi$ using a  
  complex number 
\[
\psi=\delta^{1/2}(x-q_{m_0})\exp{\rm i}\theta[q].
\]
(Since observable quantities do not involve ill defined function $\delta^{1/2}(x)$, 
 we continue to use it.)  The phase $\theta[q]$ is assumed to depend on ${\mathcal{P}}$ 
and for every $t_n$, $\psi$ is summed up by applying 
\[
\sum_{\rm time}\equiv  
 \prod_{n=0}^{\infty}(C\sum_{m_{n}=-M/2}^{M/2}).
\]  
($C$ is determined later.) Writing  
$\psi=\psi(x, [m_n ])$, the usual wave function $\Psi$ is defined by 
 \[
\Psi(x ,t)
=A\sum_{\rm time}\psi(x ,[m_n ])
=A\sum_{\rm time}\delta^{1/2}(x-q_{m_0})\exp{\rm i}\theta[q],
\]
where $A$ is the normalization factor.   
 Thanks to coarse graining, 
 one can apply the hydrodynamic expansion to 
$\theta[q]$. Keeping up to the square of the time derivative  
and assuming the time reversal invariance, 
\[
\theta[q]\sim       
   \int_{-\infty}^{t} L~\! dt/\hbar,~~~~~~L=-V(q)+m(q)~\!\dot{q}^2 /2.
\]  
 At present the $q$ dependence 
  of the mass term is not observed, so we set $m(q)=m$.  
Assuming $\sum_{m_{n}}\rightarrow \rho\int\! dq$ as $M\rightarrow\infty$ with the 
constant density $\rho$, 
 the path-integral form\cite{Feynman} is recovered 
 if $C=(1/\rho\ell_d )$ with 
$\ell_d =\sqrt{2\pi\hbar\Delta t/m}$, the diffusion length in $\Delta t$. Thus 
in each time slice, $\sum_{m_n}/\rho$ (average in unit length) 
appears, which is independent of $\rho$. 
Our sum (or average) is over mutually exclusive points   
which is equivallent to the insertion of the complete set of the coordinate.  
The particle-like factor $\delta^{1/2}(x-q_{m_0})$ has changed into the wave by 
averaging and the deternimism is masked by the large fluctuation.
 Although $q$ is just fluctuating, as expressed by the uniform sum over $q_{m_n}$,
 $\Psi(x,t)$ obeys SE and various $\Psi$'s are produced by the choice of the Hamiltonian. 
The conservation of the momentum, energy e.t.c. hold in $t$-scale in the operator form as usual. 
 The weight of each ${\mathcal{P}}$ contributing to $\Psi$ is determined by the 
Hamiltonian and in case ${\mathcal{P}}$'s connecting some points drop out of the sum 
by the phase cancellation, one can discriminate $q_{m_n}$'s by 
 separating them into coherence groups, the coherence remaining in one group. 

\vspace{0.3cm}
\begin{center}
{\bf Sample average}
\end{center}
\vspace{0.2cm}

Consider $L$ samples of $q(\tau)$, having the same initial $\Psi(x,t_I )$ 
(defined by the time average). 
Here $L$ is a large number of the same order of $M$. 
At some fixed time in $D_{t_n}$ specified by $m'_n$,  
let us denote the position of $l$-th sample 
by $q^l_{m'_n}\equiv q^{l_n}$. For every $t_n$, these 
 are summed up independently by  
 \[
\sum_{\rm sample}\equiv \prod_{n=0}^{\infty} 
(C'\sum_{l_n =1}^L ).
\]
 Here $C'=(1/\rho'\ell_d )$ with 
$\rho'$ the density of sample points which fill up the whole space. 
The phase $\theta[q^l ]$ depends on 
${\mathcal{P}} \equiv (q_{l_{0}},q_{l_{1}},q_{l_{2}},\cdots)$. 
By $\sum_{\rm sample}$, we sum up paths connecting $q^{l_n}$'s of $L$ samples in all possible ways. 
This is the coherence in the sample average, which makes the sample average equal to the time average.   
Defining
\[
\psi(x,~\![l_n ])=
 \delta^{1/2}(x-q^{l_0})\exp{\rm i}\theta~\![~\!\!q^l ],
\]
 $\Psi$ has another form; 
\begin{equation}
\Psi(x,t)=A\sum_{\rm sample}\psi (x,[~\!l_n ]).     
\label{TL}
\end{equation} 
The separation of sample points into coherence groups by the nonvanishing 
weight of ${\mathcal{P}}$'s works for macrovariables.  

\vspace{0.3cm}
\begin{center}
{\bf Macrovariables}\cite{fukuda,RFukuda}
\end{center}
\vspace{0.2cm}

For any macroscopic quantum system,
 the motion as a whole is classical. 
This is explained by the stationary phase accompanying the macrovariable $X$, defined 
 by the average of a large number 
$N$ of microscopic degrees. 
Indeed, for a thermodynamically normal system, 
the action functional has the form $NS[X]$, and suppresses the size of
 the fluctuation $d_N$ of $X$ to $O(1/\sqrt{N})$ around the stationary path 
 as is seen clearly in the path-integral formalism. 
 Non-diffusive sharp peak 
 appears in the wave function $\Psi(X,t)$\cite{fukuda}, which is the signal in the experiment. 
Take the simplest case of the center of mass of
 microcoordinates $y_i$ of mass $\mu$,
 $X=\sum_i y_i /N$. 
For infinite $N$, $|\Psi(X,t)|$ has the
 form $\delta^{1/2}(X-X^{\rm st}(t))$ with    
$X^{\rm st}(t)$ describing a smooth stationary path.  
   The Hilbert space ${\rm H}_{X^{\rm st}}$ is labeled by the continuous 
$X^{\rm st}$ and spanned by microvariables. 
 The fluctuating velocity   
\[
\hbar/({\rm i}N\mu)\partial/\partial X=
 (X(t+\Delta t)-X(t))/\Delta t
\]
becomes $\dot{X}$ evaluated 
along $X^{\rm st}$ and 
 the density or the energy density written by the quantum mechanical rule   
 becomes the classical expression;  
\[
\Psi^* (X,t)(1,P^2 /2N\mu)\Psi(X,t)~~\rightarrow~~(1,N\mu\dot{X}^{{\rm st}^2}/2)\delta(X-X^{\rm st}).
\] 
When $N$ is large but finite, 
 $d_N$ increases with $t$, but it takes 
extremely long time for $d_N$ to change its size appreciablly.   
 Two peaks with the distance larger than $d_N$
cannot be connected by $\mathcal{P}$ with the sizable weight for $\Psi$. 
 They are in different coherence groups and 
macroscopically distinguishable. 
 Our arguments below apply for any 
 $X$ as long as it loses fluctuations.  

\vspace{0.3cm}
\begin{center}
{\bf Meaurement by freezing the object state}
\end{center}
\vspace{0.2cm}

 Let $X$ be the variable of the detector, which is 
 switched on at $t_s$ to measure   
 the object operator $O$. Its eigen-states are written as        
 \[
O|a\!\!>=\lambda_a |a\!\!>, ~~~~\phi_a (x)=<\!\!x|a\!\!>.
\]  
Just before $t_s$, the wave function of the object plus detector
 $\Psi(x,X,t)$ is assumed to be factorized as 
$(\sum_a c_a \phi_a (x))\Psi^0 (X)$ with $\Psi^0 (X)$ having a peak 
  at $X=X^0$. 
The total Hamiltonian is the sum of three terms, 
 object, detector and interaction between the two;  $H=H_O +H_D +H_I $. 
Besides $H_D$, $H_I$ is $O(N)$ since the detector is arranged in such a way that 
the object interacts with a large number of micro-coordinates in the detector. 
 For definiteness, we assume $H_I = H_I (O,X)$. 
As long as the detector is on, $H_I$ is not zero, so 
$H_O$ can be neglected since it is $O(1)$. 
More precisely, one can expand for large $N$ 
in powers of off-diagonal part of $H_O$; $<\!\!b|H_O |a\!\!>,
\hspace{0.1cm} a\neq b$.  
In such a situation,
 we can use the time evolution operator
$U_a (t,t_s )$ written by $H_a \equiv H_D +
 H_I (\lambda_a ,X)$ for each $a$.  
 After $t_s$, $\Psi(x,X,t)$ becomes  
\[
\sum_a {\rm e}^{-{\rm i}\omega_a T}c_a \phi_a (x) U_a (t,t_s )
\Psi^0 (X), ~~~~~~\omega_a =<\!\!a |H_O |a\!\!>\!\!/\hbar, ~~T=t-t_s.
\] 
If $H_I \neq 0$, the object does not fluctuate between 
different states $|a\!\!>\leftrightarrow|b\!\!>$, 
during which $X$ changes its value. 
Freezing the fluctuating object is an essential mechanism 
 of the measurement by the stationary phase.   
Since the phase of $U_a (t,t_s )$ is $O(N)$ and
 produces different stationary path $X_a (t)$ for different 
$a$, the branching of the stationary phase are realized through 
the mapping $|a\!\!> \hspace{0.1cm}\!\leftrightarrow U_a
  \hspace{0.1cm}\!\leftrightarrow X_a (t)$;  
 $\Psi (x,X,t)=\sum_a c_a \Psi_a (x,X,t)\equiv\sum_a \tilde{\Psi}_a (x,X,t)$
 with $\Psi_a$ having a normalized peak at $X_a (t)$.     
 The signal function defined by 
 $J(X)\equiv \int|\Psi(x,X,t)|^2 dx$ evolves as 
\begin{equation}
  \delta (X-X^0 )\longrightarrow \sum_a |c_a |^2 \delta (X-X_a (t) ). 
\label{p1-p}
\end{equation}
for infinite $N$. 
 This is all that the ordinary quantum mechanics can tell us. 
For more general case where $H_I =\sum_i H_I (O,y_i )$ and 
$X=\sum_i f(y_i )/N\equiv X(\mbox{\boldmath $y$})$
 with some function $f(y)$,  
we can show that 
eq.(\ref{p1-p}) holds if $J(X)$ is replaced by 
\[
\int\!dx\!\int\!d\mbox{\boldmath $y$}\delta(X-X(\mbox{\boldmath $y$}))|\Psi(x,\mbox{\boldmath $y$} ,t)|^2.
\]  

\vspace{0.3cm}
\begin{center}
{\bf Structure of $\mbox{\boldmath $X$}^{\rm st}$}
\end{center}
\vspace{0.2cm}

Consider the center of mass case.
Just as $q(\tau)$ for $x$,
 let $r_i (\tau)$ corresponds to 
 $y_i$. Then $Q\equiv \sum_i r_i /N$ is the deterministic variable of $X$. 
 $\Psi(X,t)$ is written by the time average of 
$\psi(X,[m_n ])$ of a
 sample, given by 
\[
\psi(X,[m_n ])=
\delta^{1/2} (X-Q_{m_0})\exp({\rm i}\theta[Q]).
\] 
When $N\rightarrow \infty$, out of many exclusive ${\mathcal{P}}$'s, 
{\it single path of  
 $Q_{m_n}$ is selected 
by the stationary phase} which is $X^{\rm st}(t)$. 
The determinism is recovered in classical form. 
 Actually, the minimum size of the fluctuation of $Q$  
 is $O(1/N)$, since its origin is due to the individual $r_i$. Therefore, 
even when $d_N \rightarrow 0$, 
 an infinite number of $Q$'s of $O(\sqrt{N})$ are contained under $X^{\rm st}$; 
i.e. $X^{\rm st}$ has the structure. 
They are in one coherence group and  
$\Psi$ is obtainable by summing up only these points for all $t_n <t$ at
 the peak (more precisely, track of the peak), neglecting other $Q$'s.  
The branch selection is done by utilizing above fluctuations (see below). 
 In the case of sample average, including the object variable, 
we sum up
\[
\psi(x,X,[l_n])=\delta^{1/2} (x-q^{l_0})\delta^{1/2} (X-Q^{l_0 })
 \exp({\rm i}\theta[q^{l} ,Q^{l} ]).
\]   
Then the following relation holds; 
\begin{eqnarray}
\Psi(x,X,t)&=&
A'\sum_{\rm sample} \psi(x,X,[l_n ] )
\label{morl}
\\&&\Rightarrow
 \delta^{1/2}(X-X^{\rm st}(t))\Psi_{X^{\rm st}}(x.t)    
\label{always}
\end{eqnarray}
 In (\ref{morl}),
 $\sum_{\rm sample}$ can be replaced by $\rho'\int\!dq\sum_{Q\subset {\rm peak}}$ and 
 $\Rightarrow$ implies the ideal limit $L,N\rightarrow\infty$. 
$\Psi_{X^{\rm st}}(x,t)$ is normalized in $x$ and is 
 obtained by 
\[
\delta^{1/2}(x-q(t))\exp{\rm i}\theta[q,X^{\rm st}]
\]
 integrated over $q$. Thus $J(X)$ becomes the density of 
a classical point-like particle. 
 Normally, $X^{\rm st}$ is not affected by mico-degrees but
 measuring devices establish the micro-macro correlation $|a\!\!> 
 \hspace{0.1cm}\leftrightarrow X_a$ by the factor 
 $\exp{\rm i}\theta[q ,X_a ]$ with 
 $\theta[q ,X_a ]$ being $O(N)$. It is just 
 the phase factor of the path integral form of $U_a (t,t_s )$. 

\vspace{0.3cm}
\begin{center}
{\bf Branch selection}
\end{center}
\vspace{0.2cm}

One sample cannot sustain two or more stationary paths
 since $Q$ has a definite position and the size of the 
 fluctuation is $O(1/\sqrt{N})$ in $t$-scale. 
Thus one sample selects one branch 
 and after that it remains in the same branch 
 unless an extra force-product (or the action) of $O(N)$ is supplied. 
 The branch selction is done by chance through 
 fluctuations of $Q$'s of the size smaller than $d_N$ 
 in the branching region $B$, see the Figure. 
The mechnism works 
 even for infinite $N$ 
because of the above stated structure of $X^{\rm st}$. 
After $t_s$, each branch constitutes a coherence group.
Now prepair $L(=\sum_a L_a )$ samples having 
$\Psi^0$ initially 
and suppose $L_a$ samples actually selected $a$-th branch.  
By writing $\Psi$ by the sample average, 
\begin{equation}
  L_a /L \Rightarrow |c_a |^2 .   
\label{PPLL}
\end{equation}
 is shown to hold. $B$ is assumed to be 
 $(t_s ,t_s +k\Delta t)$ with finite $k$, becoming  
 $(t_{B-},t_{B+})$ as $\Delta t\rightarrow 0$. 
Indeed, the branch selection occurs instantaneously at $t_s =t_{B\pm}$, since 
a new stationary phase is created by  
 $H_I (\lambda^a ,X)$, leading to the phase of the form
 $N\int_{t_s}^t 
{\mathcal{L}}(\lambda^a ,X;t)dt$.
 Thus $t-t_s$ can be infinitesimal for infinite $N$. 
\begin{figure}
\begin{center}
\includegraphics{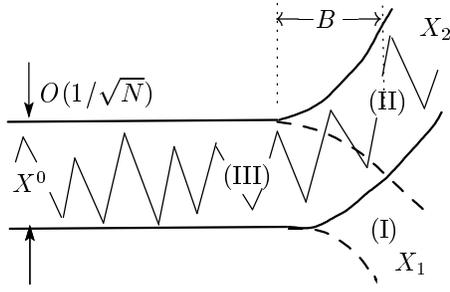}
\end{center}
\begin{center}
\caption{\label{fig1:epsart}
 Schematic behavior of $Q$
 choosing the branch $X_2 $ in two branch case. $\kappa_a $ 
 in $({\rm (I),(II),(III)})$ is 
 $(0,1,L_2 /L)$. Effective $\kappa_a$ in $B$ is $\sqrt{L_2 /L}$.
 The figure is for large but finite $N$.}
\end{center}
\end{figure}
Assuming a perfect mapping $|a\!\!>\leftrightarrow X_a$, 
one has only to concentrate on $Q$'s by writing 
$\sum_{\rm sample}=\rho'\int\!dq\sum_{Q\subset {\rm peak}}$. 
Before proving (\ref{PPLL}), 
 the density of $Q$'s of $a$-th group and its effect on 
the state vector are studied. 

\vspace{0.3cm}
\begin{center}
{\bf Density of points and coherence}
\end{center}
\vspace{0.2cm}

 Prepair sufficient number of samples,
 all having the initial wave function $\Psi^0$ and
 assume that their $Q_{m'_n}^l$'s
 with arbitrarily chosen $m'_n$ for every time slice $t_n$
 fill up the whole space with the constant density denoted by $\rho_L$
 and hence at the peak position also. 
$\Psi$ defined by the sample average of them   
 satisfies SE, becoming $\sum_a c_a \Psi_a$ after $t_{B+}$. 
 The number of samples to be summed over is $L=\rho_L d_N =O(\sqrt{N})$. 
 Consider $\sum_{l_n \subset {\rm peak}}/\rho_L$ appearing
 for all $t_n <t$ in the path integral.    
Suppose $t>t_{B+}$ and $L_a$ out of $L$ selected $a$-th branch.
Then, for $t_n <t_{B-}$, $L$ points of $Q$'s 
make up the peak at $X^0$ with the density $\rho_L$
 and for $t_n >t_{B+}$,
 $L_a$'s constitute $X_a$ with the density also $\rho_L$. 
 Define $\kappa_a =\rho_a /\rho_L$ with $\rho_a$ the density of $a$-th group. 
When $t_n <t_{B-}$, the detector is not yet switched on so there is  
 no group difference.
 Hence $\kappa_a =L_a /L$ and $\Psi$ is obtained
 by applying to $\psi^l$ either $\sum_{{\rm all}~\!L}/\rho_L$
 or $\sum_{l\subset L_a}/(\kappa_a \rho_L)\equiv K_a$. 
Both become $\Rightarrow \int_{X^0} dQ$ (the  
integration being done around $X^0$), and 
produce the same normalized $\Psi$. 
In general, the wave function of one sample $\Psi^{(l)}$, 
if it selected $a$-th branch, is obtained by applying $K_a$
 for any $t_n <t$, provided we have 
${\mathcal{P}}'s$ connecting $Q$'s within $a$-th group in all possible ways. 
Indeed, for all $t_n$, $K_a \Rightarrow \int dQ$, which is 
the correct path integral measure and since the sum is over $a$-th group only,
 the selected branch $X_a$ is reproduced. Note that $\Psi^{(l)}$ is 
independent of $\kappa_a$. 
When $t_n >t_{B+}$, 
$\kappa_a \!=\!1$ at the peak position $X_a$ as stated above.  
The result $\kappa_a =1$ also follows from (\ref{p1-p});
 if $\kappa_a \!\!<\! 1$, a factor $\kappa_a$ appears 
 for each $t_n$ of $a$-th branch with $t_n \!\!>\!\!t_{B+}$,  
 contradicting with (\ref{p1-p}). 
It looks as if $Q^{l_n}$'s of one group
 ``gather together while fluctuating" at the exit of $B$, as in the Figure. 
Such a picture precisely coincides with the fact that 
in order to change the value of 
$Q$, the fluctuation of individual $r_i$ has to be correlated as a whole.  
 This is realized by $H_I (O,X)$. 
Intuitively, the width of the peak is reduced, which will be seen to be the case. 
 $L_a /L\leq\!\!\kappa_a \!\leq\!\!1$ 
  represents the degree of (in)coherence among different groups. 
Finally, for $t_n =t_{B\pm}$, 
apart from $K_a \Rightarrow\int dQ$, 
no time evolution factor comes in 
since the branch selection is instantaneous. All these are rederived  
using the state vector below. 
Summarizing, $\Psi^{(l)}$ defined by $K_a$  
 equals to $\Psi_a (x,X,t)$, given by (\ref{always}) with
 $X^{\rm st}(t)=X^0$ for $t<t_{B-}$ and $X^{\rm st}(t)=X_a (t)$ for $t>t_{B+}$. 
 In contrast to $\Psi^{(l)}$, 
 $\Psi$ carries a factor $\kappa_a$ coming from $B$, which 
 turns out to be $\sqrt{L_a /L}$ ---
 half coherence in $B$. 

\vspace{0.3cm}
\begin{center}
{\bf State vector}
\end{center}
\vspace{0.2cm}

Writing $Q^l$ as $X^l$, the sample sum at some $t_n $ can be replaced by the 
 equivallent insertion of the complete set $|x,X^{l}\!\!>$ at $t_n$,
 which satisfies 
\[
<\!x,X^l |x',X^{l'}\!\!>=\!\delta(x-x')\delta_{l,l'}.
\] 
Suppose $t>t_{B+}$ and define for $t_n <t_{B-}$,   
 \[
P^l \!\equiv\!\int\!dx|x,X^l \!\!><\!\!x,X^l |.
\]  
 Since $\sum_{l;{\rm all}L~\!}\Rightarrow \rho_L \!\int_{X^0}\!dX$, 
 usual relations in the continuum picture emerge      
if one identifies 
$\sqrt{\rho_L}|x,X^l \!\! >\hspace{0.1cm} \Rightarrow \!|x,X \!\!>$.  
Indeed, 
\[
\sum_{l;{\rm all}L~\!} P^l \Rightarrow 
\!\!\int\!dx\int_{X^0}\!dX|x,X \!\!><\!\!x ,X |=I^0,
\] 
where $I^0$ is the identity operator at the peak $X^0$.
Consider then, 
$P_a \!=\!\sum_{l\subset L_a}P^l $. 
Since $\kappa_a =L_a /L$, we get 
\[
P_a \hspace{0.1cm}\Rightarrow(L_a /L)\int\!dx\int_{X^0}\!dX |x,X\!\!><\!\!x,X|
\equiv (L_a /L)~\!I^0 . 
\]
 One can ascribe the factor $L_a /L$ to the state vector  
by defining $\sqrt{\rho_L}|x,X^l \!\!>\hspace{0.1cm}\Rightarrow\!\sqrt{L_a /L} |x,X\!\!>$ 
(phase is irrelevant for subsequent discussions), with 
$\int\!dx\int\!dX$ as the integration measure.
 Newly defined $|x,X\!\!>$ is the usual state vector which 
does not distinguish the group and is used in what follows.
 As stated, $X^{\rm st}$, hence $|x,X>$ has the structure
 which is represented by the factor $\sqrt{L_a /L}$.  
 When $t_n >t_{B+}$, thanks to $\kappa_a \!=\!1$,
 \[
P_a \hspace{0.1cm}\Rightarrow\hspace{0.1cm}
\int\!dx\int_{X_a}\!dX|x,X\!\!><\!\!x,X|=I_a .    
\]
 Here $I_a$ is the identity operator at $X_a$. 
Summing over $a$, one gets the evolution $I^0 \!=\!\sum_a (L_a /L)I^0 
  \rightarrow\sum_a I_a$ and the Hilbert space branches as   
  ${\rm H}_{X^0}\!=\!\sum_a (L_a /L){\rm H}_{X^0}
 \rightarrow\sum_a {\rm H}_{X_a}$ according to the density $L_a /L$.

\vspace{0.3cm}
\begin{center}
{\bf Derivation of (\ref{PPLL})}
\end{center}
\vspace{0.2cm}

 Consider the usual time evolution
\[
|\Psi\!\!> =U(t,t_{s} )U^0 (t_{s} ,t_I )|\Psi\!\!>_I
\]
with $t_s =t_{B\pm}$
 and $t_I <t_s$. 
 Here $U(t,t_{B+})$ is written by $H_D +H_I (O,X)$. 
Then $<\!\!x,X|\Psi\!\!>=\sum_a c_a \Psi_a (x,X,t)$.
 Let us write $\Psi$ by the sample average. 
 Regarding $B$ as a black box, introduce the branch selection operator 
\[
P^B_a \!=\!\sum_{l\subset L_a}
\!\int\!dx|x ,X^l_{+}\!\!><\!\!x ,X^l_{-} |.
\] 
 Here $X^l_{\pm}$ denote $X^l$ at $t_{B \pm}$  
 and separately for $\pm$, $<\!\!x,X^l_{\pm}|$'s are ortho-normal. 
$P^B_a$ maps $X^l_{-}$ to $X^l_{+}$ 
 across $B$. Now, consider
\[
|\Psi\!\!>_a =U(t,t_{B+})P^B_a U^0 (t_{B-}\hspace{0.1cm},t_I )|\Psi\!\!>.
\] 
 Writing  
 $X^l_{\pm}\Rightarrow X_{a\pm}$ if $l\subset L_a$, we see from above results that   
 \[
\sqrt{\rho_L}|x,X^l_{-}\!\!>\hspace{0.1cm}
\Rightarrow\sqrt{L_a/L}|x,X_{a-}\!\!>,~~~~~\sqrt{\rho_L}|x,X^l_{+}\!\!>\hspace{0.1cm}\Rightarrow~\!|x,X_{a+}\!\!>.
\]  
Replacing $U(t,t_{B+})$ by $U_a (t,t_{B+})$ by the assumed perfect mapping
 $|a>\leftrightarrow X_a$, we see that
 \[
U(t,t_{B+})P^{B}_a \!\Rightarrow U_a (t,t_{B+})
\sqrt{L_a/L}\int\!dx\int\!dX|x,X_{a+}\!\!><\!\!x,X_{a-}|.
\]  
When we compare this expression with the usual $\Psi$, we set at this point  
$X_{a\pm}=X^0$. 
Thus $P^{B}_a \!\Rightarrow \sqrt{L_a /L}~\!I^0$, and 
$\sqrt{L_a /L}$ is identified with the effective $\kappa_a$ in $B$. 
Apart from the factor $\sqrt{L_a /L}$,
 the unitary time evolution by $U_a U^0$ is assured, 
which is equal to $\Psi_a $. Therefore, 
$<\!\!x,X|\Psi\!\!>_a 
=\sqrt{L_a/L}\Psi_a (x,X,t)$. 
When one sums over $a$, all possibities are exhausted and ordinary $\Psi$ is reproduced 
by (\ref{morl}).   
In this way, $\sum_a \sqrt{L_a /L}\Psi_a \Rightarrow\sum_a c_a \Psi_a$
 holds and $\sqrt{L_a /L}\Rightarrow c_a$ is obtained, since 
each term has a peak at a distinct position. Thus (\ref{PPLL}) is proved.
More directly, since the mapping between $a$-th group and the branch 
$X_a$ is one to one, we can set $\sqrt{L_a /L}\Psi_a =\tilde{\Psi}_a =c_a \Psi_a$.  
 Then (\ref{PPLL}) follows for each $a$.
 If one considers $U(t,t_P )P_a U(t_P ,t_I )|\Psi\!\!>$,   
its norm squared is $(L_a /L,~\!|c_a |^2 )$
 for $(t_P \!<\!t_{B-},~\!t_P \!\!>\!t_{B+} )$. 
 By (\ref{PPLL}), the sub-norm defined above is invariant for $P_a$ 
inserted at any time $t_P$. One can show that it corresponds to the
 conservation of the number of 
points of $Q$'s along the selected channel, which is $L_a$. Thus the 
number conservation again leads to (\ref{PPLL}). 
Writing (\ref{PPLL}) as $(L_a /L)\times 1 =
1\times |c_a |^2$, and remembering that  
 the number 
$L_a$ =(density)$\times$(width),   
 the ratio of the width 
of the peak $X_a$ and $X^0$ turns out to be $|c_a |^2$, which  
explains the picture of ``gathering together'' in $B$.     
In case $X \!=X(\mbox{\boldmath $y$})$, 
 we sum up   
$\int\!dx|x,\mbox{\boldmath $y$}^l_{+}\!\!><\!\!x,\mbox{\boldmath $y$}^l_{-}|$ over $\mbox{\boldmath $y$}^l$ 
by fixing $X(\mbox{\boldmath $y$}^l )$ to $X^l$. 
 Then 
 eq.(\ref{PPLL}) can be shown.  

\vspace{0.3cm}
\begin{center}
{\bf Time average}
\end{center}
\vspace{0.2cm}

Take a sample having $\Psi^0 (x)$ at some $t<t_{B-}$.  
 Let $M$ be the number of points in the time average case, 
i.e. the number of times of fluctuations in $d_N$.  
For simplicity, we take $M=L$ and suppose $t>t_{B+}$.
  For any time slices $t_n <t$, let us define $M_a =L_a $ in such a way
 that $(M, M_a )$ consist of the same sample 
points of (whole $L$, $a$-th group). 
Now, the usual $\Psi$ is obtained by summming up $M$ points,
 i.e. summig up over $a$,  
 for any time slices, including the branching region.
 Then 
$\Psi$ becomes equal to that obtained by summing all $L$ samples, 
resulting in the linear combination of various branches, even for one sample.   
Now $M_a /M\!=\!L_a /L$,  
can be regarded as the fraction of time one sample 
stays in $a$-th branch at $t_{B-}$. 
The branching ratio is proportional to this quantity. 
 By similar arguments as above, 
$\Psi=\sum_a \sqrt{M_a /M}\Psi_a$ holds.  
  Actual $\Psi^{(l)}$ for $l$-th sample, if it selects $a$-th branch,
 is obtained by summming up $M_a$ points only,   
 divided by $(M_a /M,\sqrt{M_a /M},1)$ for $(t_n <t_{B-},t_n =t_{B\pm} ,t_n >t_{B+})$. 

\vspace{0.3cm}
\begin{center}
{\bf Discussions}
\end{center}
\vspace{0.2cm}

In our theory, all sorts of irreversibility in the detection process stem from 
the complete construction and destruction of the coherence 
among $X_a$'s by the stationary phase 
which cannot be restored by any means. 
The superposition of 
macroscopically distinct states appears since $\Psi$ is defined 
 by summming up all mutually exclusive 
points of $Q$'s in $B$, without 
discriminating the definite positions in $\tau$-scale. 
 This is the  
intepretation by our language of
 the many world picture\cite{Everett,Wheeler,Hemno} and
 that of Copenhagen.  
Since our theory is based on the quantum mechanics of $N$ particle system, 
the correction to the results obtained here 
is calculable in the form of $1/N$ expansion. 
 In contrast to \cite{Rimini,ABassi}, the 
extension to the relativistic case is quite natural if we set  
$t$ and $x$ on equal footing and adopt the field theory.

\end{document}